\documentclass[12pt]{article}

\usepackage[dvips]{graphicx}

\usepackage{amsmath}
\usepackage{amsfonts}
\usepackage{amssymb}
\usepackage{cite}
 
\usepackage[a4paper]{typearea} 
  \areaset[0cm]
          {17.0cm}{24cm}       

\newcommand{\BILD}[4]{\begin{figure}[#1]%

     #2

     \centerline{\parbox{15cm}{\caption[.]{#3} \label{#4}}}
     \end{figure} }

\newcommand{\Vektor}[2]{\begin{pmatrix} #1 \\ #2 \end{pmatrix}}

\newcommand{\R}{\mathbb{R}}
\newcommand{\Z}{\mathbb{Z}}
\newcommand{\N}{\mathbb{N}}

\newcommand{\T}{\mathbb{T}}

\newcommand{\Int}{\int\limits}

\newcommand{\e}{\operatorname{e}}

\newcommand{\ud}{\text{d}}
\newcommand{\ui}{\text{i}}

\newcommand{\cN}{{\cal N}}

\newcommand{\setsep}{ \;\; | \;\;}

\newcommand{\eqnref}[1]{eq.\ \eqref{#1}}

\newcommand{\tmod}{\;\;\text{mod}\;}

\newcommand{\uexp}[1]{\exp\left(#1\right)}

\newcommand{\defin}{\mathrel{\mathrm{\raisebox{.095ex}{:}\hspace*{-0.9ex}=}}}
\DeclareMathOperator{\Tr}{Tr}

\begin{document}


ULM--TP/99-3

May 1999

\vspace*{2cm}

\newcommand{\titel}{Spectral statistics for quantized \\[0.5ex] 
                    skew translations on the torus}

\normalsize

\vspace{0.5cm}

\renewcommand{\thefootnote}{\fnsymbol{footnote}}

\begin{center}  \huge  \bf

      \titel
\end{center}

\setcounter{footnote}{1}

\begin{center}
   \vspace{3cm}
 
         {\large Arnd B\"acker%
\footnote[1]{E-mail address: {\tt arnd.baecker@physik.uni-ulm.de}}
          and Grischa Haag%
\footnote[7]{E-mail address: {\tt grischa.haag@physik.uni-ulm.de}}
          }

   \vspace{4ex}

   Abteilung Theoretische Physik\\
   Universit\"at Ulm\\
   Albert-Einstein-Allee 11\\
   D--89069 Ulm\\
   Federal Republic of Germany\\

\end{center}

\vspace*{2.5cm}

\noindent
{\bf Abstract:}

\noindent
We study the spectral statistics for quantized skew translations
on the torus, which are ergodic but not mixing for irrational 
parameters. It is shown explicitly that 
in this case the level--spacing distribution and other common 
spectral statistics, like the number variance,
do not exist in the semiclassical limit.

\renewcommand{\thefootnote}{\arabic{footnote}}
\setcounter{footnote}{0}


\newpage

\section{Introduction}

One of the central questions in quantum chaos is how the
asymptotic distribution of the energy levels of a quantum system depends 
on the behaviour of the corresponding
classical dynamical system.
For integrable systems the spectral statistics have been
conjectured \cite{BerTab77} to be Poissonian,
whereas chaotic systems have been conjectured \cite{BohGiaSch84}
to be described by random matrix theory (like
the Gaussian orthogonal ensemble for systems with time--reversal
symmetry).
Both conjectures are supported by many numerical studies.
However, in both cases exceptions are known:
so-called arithmetic systems (see e.g.\
\cite{AurSte88,BogGeoGiaSch92,BolSteSte92,Bol93,Sar95})
show Poissonian spectral statistics
despite being strongly chaotic. 
Another example showing non--generic spectral statistics 
are quantized cat maps \cite{HanBer80,Kea91c}.
As an example for a class of integrable systems 
the eigenvalues statistics for flat tori are studied in \cite{Sar97}.
It is proven that the pair correlation function is Poissonian
for a set of full Lebesgue measure in the parameter space of tori,
but that it does not exist for a set of second Baire category
(a topologically large set). Explicit examples of tori
with Poissonian pair correlation function are given in \cite{EskMarMoz98:p}.
A further class of integrable 
systems showing exceptional behaviour are harmonic oscillators,
for which the nearest neighbour level--spacing and other spectral
statistics do not possess a limit distribution,
see e.g.\ \cite{BerTab77,PanBohGia89,Ble90,Ble91,Gre96,Mar98:p} 
and references therein. 

An important class of model systems for studies in quantum chaos
arise from the quantization of area preserving maps, 
see e.g.~\cite{HanBer80,BerBalTabVor79} and references therein.
In this paper we study the spectral statistics, i.e.\ the
distribution of eigenphases, for the class of quantized skew translations
on the torus (also called parabolic maps) 
\cite{BouBie96,DeBDegGia96,MarRud99:p}.

\section{Spectral statistics}

A particular example of a 
skew translation on the torus $\T^2$ (see e.g.\ \cite{CorFomSin82}) is defined by 
\begin{equation}
\Vektor{p}{q} \xrightarrow{ A_\alpha} \Vektor{p+\alpha}{q+2p} \tmod 1 \;\;, 
\end{equation}
where $\alpha\in\R^+$ determines the dynamical behaviour:
For rational $\alpha$ the mapping is not ergodic, whereas for irrational
$\alpha$ the map is ergodic
and, in particular,  uniquely ergodic \cite{Fur61}, 
i.e.\ there is only one invariant
ergodic measure, a situation rarely encountered for a dynamical system.
This implies that $A_\alpha$ does not possess any periodic
points for $\alpha$ irrational. Moreover $A_\alpha$ is not mixing, see 
e.g.\ \cite{CorFomSin82}.

A quantization of an area--preserving map on the torus
is given by a sequence of unitary time evolution operators $U_N$ defined
on an $N$--dimensional Hilbert space,
where $N\to\infty$ corresponds to the semiclassical limit.
As quantization 
 of these skew translations
we use the one proposed in \cite{MarRud99:p},
which is based on considering appropriate rational
approximations $a_N/N$ to $\alpha$.
That is, for a given irrational $\alpha$ and $N\in\N$
there is a unique $a_N\in\N$ defined by the condition
\begin{equation} \label{eq:cond-a_N}
  \left| \alpha -\frac{a_N}{N} \right| < \frac{1}{2N} \;\;.
\end{equation}
Then the propagator $U_N$ 
can be expressed in the position representation 
by the $N\times N$ unitary matrix
\begin{equation}
  (U_N)_{kj}= \frac{1}{N}\sum_{l=0}^{N-1}
  \uexp{\frac{2\pi\ui}{N}\left(lk-(l-a_N)^2-(l-a_N)j\right)}\;\;.
\end{equation}
with $j,k\in\{0,1,\dots,N-1\}$.
One investigates the eigenvalues $\e^{\frac{2\pi\ui}{N}\phi_j}$ of $U_N$,
where
$\phi_j\in[0,N[$ and $j\in\{0,\dots,N-1\}$.
The spectral density $\varrho(\phi)$ is given by 
\begin{equation} 
  \label{eq:time-evolution-propagator}
  \varrho(\phi)
  \defin\sum_{j=0}^{N-1}\sum_{k\in\Z}\delta
       \left(\tfrac{2\pi}{N}(\phi-\phi_j)-2\pi k\right) 
  \;\;,
\end{equation}
and using the Poisson summation formula $\varrho(\phi)$ can be 
expressed in terms of $U_N$ by
\begin{equation} 
  \label{eq:time-evolution-propagator-powers-U}
  \varrho(\phi)
  =\frac{1}{2\pi}\sum_{l\in\Z} \e^{\frac{2\pi\ui}{N}l\phi} \Tr U^l_N
  \;\;.
\end{equation}
For the skew translations the eigenphases of the matrix $U_N$ can be
determined explicitly \cite{MarRud99:p}
\begin{equation} \label{eq:ev}
   \phi_{\eta,l} = l D - \eta^2 + \eta a_N - a_N^2 \frac{(M-1)(2M-1)}{6}
   \tmod N 
  \;\; 
\end{equation}
with $\eta\in\{1,\dots,
D\}$, $l\in\{0,\dots,M-1\}$ and $M=N/D$, where $D=\gcd(a_N,N)$ is the 
greatest common divisor of $a_N$ and $N$.

An important statistics is the level--spacing distribution, which
is the probability density for the distribution 
of the distances $\phi_{j+1}-\phi_j$ between (unfolded) eigenphases 
$\phi_j \in [0,N[$.
More precisely, one considers (with $\phi_N \defin \phi_0$)
\begin{equation}
  \lim_{N\to\infty} 
    \frac{\#\{   j < N \setsep a \le  \phi_{j+1}-\phi_j \le b \} }{N}
    = \Int_a^b P(s) \; \ud s \;\;
\end{equation}
if a limit distributions $P(s)$ exisits.
From \eqnref{eq:ev} follows $\phi_{\eta,l}+D=\phi_{\eta,l+1}$ and 
consequently the spectrum is periodic with period $D$.
Moreover, the last term in \eqnref{eq:ev} is independent of $\eta$ 
and $l$ such that for the level--spacing  distribution  
it is sufficient to study the reduced spectrum
\begin{equation}
  \label{eq:ev-red}
  \varphi_\eta
  \defin - \eta^2 + \eta a_N \tmod D 
  = -\eta^2 \tmod D 
\end{equation}
with $\eta \in\{1,\dots,D\}$. For a fixed $\alpha\in\R^+$ and a 
given $N$ \eqnref{eq:cond-a_N} fixes a 
rational approximant $a_N \in \N$ and also $D=\gcd(a_N,N)$. 
Let us consider three special cases. 
First assume that $D=1$. Then the reduced spectrum \eqnref{eq:ev-red}
consists of just one number, i.e.\ the original spectrum \eqnref{eq:ev} is 
completely rigid, leading to a level--spacing distribution 
\begin{equation}\label{eq:level-spacing-D-one}
  P_{D=1}(s)=\delta(s-1) \;\;.
\end{equation}
Assuming $D=2$ we get for the reduced spectrum
\begin{equation}
  \label{D-two}
  \varphi_1=-1 \tmod 2 \equiv 1
  \qquad\text{and}\qquad
  \varphi_2=-4 \tmod 2 \equiv 0
  \;\;.
\end{equation}
Thus the spectrum \eqnref{eq:ev} is composed of
two subsequences $\phi_{1,l}$, $\phi_{2,l}$ with an equidistant 
spacing of $D=2$. Since these two subsequences are shifted with respect to 
each other by $3 \tmod 2\equiv 1$, we obtain for 
the level spacing distribution $P_{D=2}(s)=\delta(s-1)$ as in the case $D=1$.
Finally we consider  the special case $D=3$. 
The reduced spectrum is given by
\begin{equation}
  \label{D-three}
  \varphi_1=-1 \tmod 3 \equiv 2
  \;,\quad
  \varphi_2=-4 \tmod 3 \equiv 2
  \qquad\text{and}\qquad
  \varphi_3=-9 \tmod 3 \equiv 0
  \;\;.
\end{equation}
Thus the spectrum \eqnref{eq:ev} consists of three subsequences. 
Two of them, $\phi_{1,l}$ and $\phi_{2,l}$, lead to the same eigenphases,
i.e.\ the spacing between them is zero. 
The spacing between these two subsequences and the third subsequence 
is $2$ and $1$, respectively. Thus we get for the level spacing distribution
\begin{equation} \label{eq:level-spacing-D-three}
  P_{D=3}(s)=\frac{1}{3} \left[ \delta(s) + \delta(s-1) +\delta(s-2) \right]
  \;\;.
\end{equation}  
Using the cases of $D=1$ and $D=3$ we show that there
is no limit distribution of the level spacing distribution
for the quantized skew translations in the limit $N\to\infty$
by an explicit construction of two different limit
points of the sequence of level spacing distributions.

A general result from the approximation theory of irrational numbers, 
see e.g.\ \cite{HarWri79}, asserts that for any 
irrational $\alpha$ there exists an infinite 
sequence of pairs $(a_N,N)$ with $a_N,N\in\N$ 
and $\gcd(a_N,N)=1$ such that
\begin{equation}
  \label{eq:irr-approx}
  \left| \alpha - \frac{a_N}{N} \right| < \frac{1}{N^2} \;\;.
\end{equation}
All these pairs are approximations fulfilling \eqnref{eq:cond-a_N}.
If $(a_N,N)$ is such an approximation
then $(a_{N'},N')=(D'a_N,D'N)$ for $N\geq 2D'$ is also 
a good approximation. This follows from
\begin{equation}
  \label{eq:approx-D}
  \left|\alpha-\frac{a_{N'}}{N'}\right|
  =\left|\alpha-\frac{a_N}{N}\right|
  <\frac{1}{N^2}
  \leq\frac{1}{2 D'N}
  =\frac{1}{2 N'}
  \;\;.
\end{equation}
This implies that for each $D\in\N$ there is an infinite 
sequence of pairs $(a_N,N)$ with $D=\gcd(a_N,N)$  
fulfilling \eqnref{eq:cond-a_N}.
With the explicit calculation of the level--spacing distribution $P(s)$ 
for $D=1$ and $D=3$ we obtain two infinite sequences 
for which the level--spacing distributions are different.
Consequently there is no limit of the level--spacing 
distribution as $N\to\infty$.

Another commonly used statistics is the number variance
which measures long range correlations in the spectrum.
For quantized maps with unfolded eigenphases $\phi_j \in [0,N[$
the number variance is defined by
\begin{equation}  \label{eq:nv-def}
  \Sigma^2(L;N) \defin \frac{1}{N} \Int_0^{N} 
          \left( \cN(\phi+L) -\cN(\phi) - L \right)^2 
          \; \ud \phi \;\;,
\end{equation}
where $\cN(\phi) \defin \int_0^\phi \varrho(\phi') \; \ud \phi'$
is the integrated spectral density.
Notice that for $L\le N$ we have $\Sigma^2(L,N) = \Sigma^2(N-L,N)$.

\BILD{b}
   {
     \includegraphics[width=\textwidth]{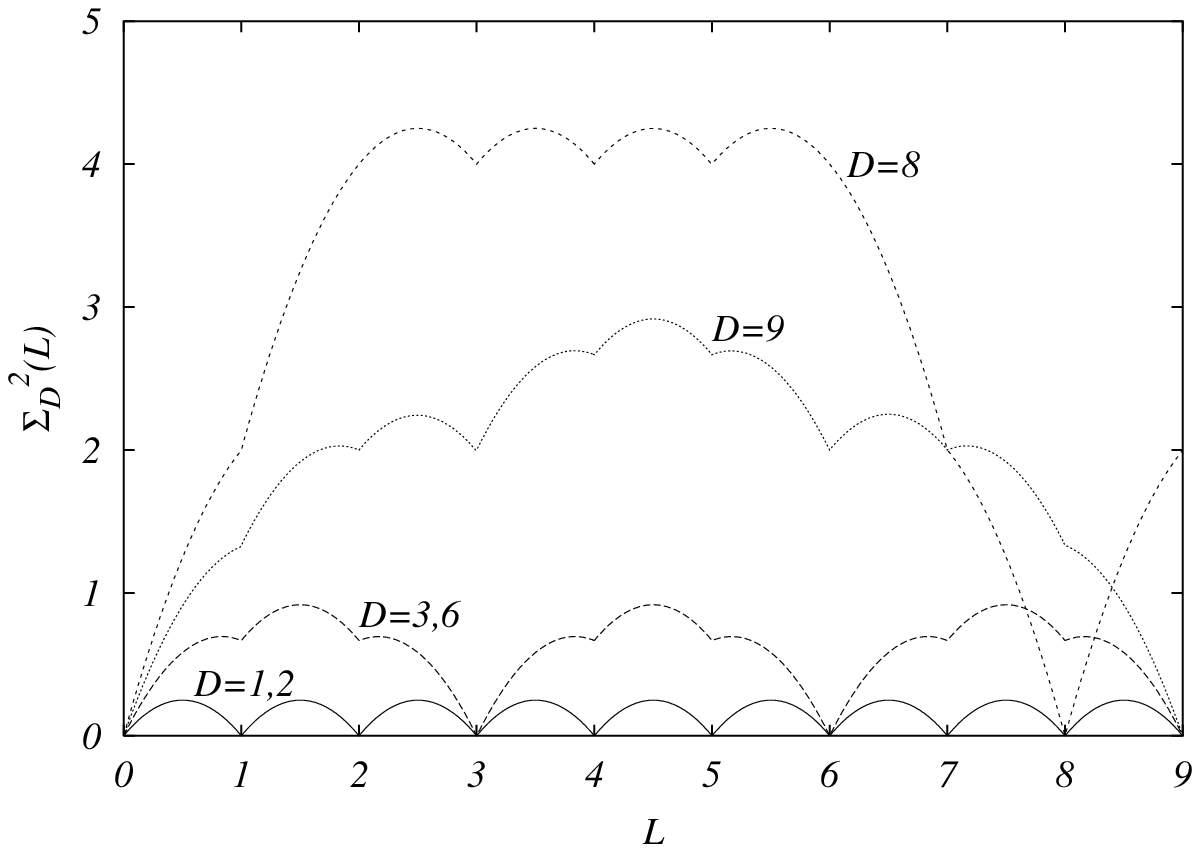}
   }
   {Number variance for the quantized skew translations on the torus
    for $D=1,2,3,6,8$ and $D=9$.}
   {fig:number-variance-skew} 

Using \eqnref{eq:time-evolution-propagator}
the number variance can be expressed 
in terms of the propagator $U_N$ 
\begin{equation}  \label{eq:nv}
  \Sigma^2(L;N) =\tfrac{2}{\pi^2}\sum_{n=1}^\infty 
     \tfrac{1}{n^2}\sin^2\left(\tfrac{n\pi L}{N}\right)
      \left|\Tr U^n_N \right|^2
     \;\;.
\end{equation}
From the explicit expression \eqnref{eq:ev} for the eigenphases 
one obtains 
\begin{equation}
  \label{eq:trace}
  \begin{split}
  \Tr U_N^n &= \sum_{\eta=1}^D \sum_{l=0}^{M-1} \uexp{n \phi_{\eta,l}}\\
  &=\sum_{\eta=1}^D \sum_{l=0}^{M-1} \uexp{\tfrac{2\pi\ui}{N}n
      \left(l D - \eta^2 + \eta a_N - a_N^2 \tfrac{(M-1)(2M-1)}{6}\right)} \\
  &=M\delta_{(n\!\!\tmod\!\!M), 0}\sum_{\eta=1}^{D} \uexp{\tfrac{2\pi\ui}{N}n
      \left( - \eta^2 + \eta a_N - a_N^2 \tfrac{(M-1)(2M-1)}{6}\right)} 
      \;\;.
  \end{split} 
\end{equation}
This implies for the number variance of quantized skew maps 
with $D=\gcd(a_N,N)$ 
\begin{equation}
  \label{eq:nv-skew}
  \Sigma^2_D(L)
  =\tfrac{2}{\pi^2}\sum_{k=1}^\infty \tfrac{1}{k^2} \sin^2 
         \left(\tfrac{k\pi L}{D}\right)\left|\sum_{\eta=1}^{D} 
                \uexp{-\tfrac{2\pi\ui}{D}k \eta^2} \right|^2
       \;\;.
\end{equation}
Notice, that $\Sigma^2_D(L)$ does not depend explicitly on $a_N$ and $N$, 
but only on their greatest common divisor $D$.
For $D=1$ we get
\begin{equation}
  \label{eq:nv-D-one}
  \Sigma^2_{D=1}(L)
  =\tfrac{2}{\pi^2}\sum_{k=1}^\infty 
             \tfrac{1}{k^2}\sin^2\left({k\pi L}\right) 
  =\big(L-\lfloor L \rfloor\big)+\big(L-\lfloor L \rfloor\big)^2 \;\;,
\end{equation}
where $\lfloor x \rfloor$ denotes the integer part of $x$.
The same result also holds for $D=2$.
In the case of $D=3$ the computation of the Fourier series involved 
leads to
\begin{equation}
  \label{eq:nv-D-three}
  \Sigma^2_{D=3}(L;N)
  = -\tfrac{8}{9}+5 F\left(\tfrac{L}{3}\right)
                +2 F\left(\tfrac{L-2}{3}\right)
                +2 F\left(\tfrac{L+2}{3}\right) \;\;,
\end{equation}
where we defined 
$F(x) \defin x-\lfloor x \rfloor + \left( x-\lfloor x \rfloor \right)^2$.
Thus the number variance is different for $D=1$ and $D=3$,
and consequently there is also no limit of the number variance
as $N\to\infty$.

In fig.~\ref{fig:number-variance-skew} we show four examples
for the behaviour of the number variance in dependence
on $D$. The cases $D=2$ and $D=6$ coincide
with $D=1$ and $D=3$ respectively, which illustrates
that $D$ is not necessarily the smallest period of $\Sigma^2_D(L;D)$.
A higher number of degeneracies in the reduced spectrum \eqnref{eq:ev-red},
as for example in the case of $D=8$, leads to large values
of the number variance.

\section{Discussion}

The non-existence of limit distributions 
for the spectral statistics of quantized skew translations
provides another counterexample to the universality 
of energy level statistics observed in many situations.
In contrast to the case of flat tori one has for
the class of quantized skew transformations explicit
examples for which the spectral statistics do not exist.
There are different possibilities to interpret this
result. 
On the one hand, this example may be seen as 
an indication that in order to obtain
the expected random matrix behaviour not just ergodicity but also
the mixing property of the classical system is needed.
On the other hand, one may consider this class of systems
as being quite non--generic, in a similar manner
as the quantized cat maps.
Finally, we would like to remark that it
may be possible that certain
spectral statistics exist for $N\to\infty$ when one 
averages over a (possibly increasing) range of
different $N$, as it has been shown for the quantized 
cat maps \cite{Kea91c}.

\vspace*{0.5cm}

\noindent
{\bf Acknowledgments}

\noindent
We would like to thank Prof.~Dr.~F.~Steiner and Roman Schubert for useful 
discussions and comments. 
A.B.\ acknowledges support by the 
Deutsche Forschungsgemeinschaft under contract No. DFG-Ste 241/7-3.



\begin{thebibliography}{10}

\bibitem{BerTab77}
M.~V. Berry and M.~Tabor: {\em Level clustering in the regular spectrum\/},
  Proc. R. Soc. London Ser. A {\bf 356} (1977) ~375--394.

\bibitem{BohGiaSch84}
O.~Bohigas, M.-J. Giannoni and C.~Schmit: {\em Characterization of chaotic
  quantum spectra and universality of level fluctuation laws\/}, Phys. Rev.
  Lett. {\bf 52} (1984) ~1--4.

\bibitem{AurSte88}
R.~Aurich and F.~Steiner: {\em On the periodic orbits of a strongly chaotic
  system\/}, Physica D {\bf 32} (1988) ~451--460.

\bibitem{BogGeoGiaSch92}
E.~B. Bogomolny, B.~Georgeot, M.-J. Giannoni and C.~Schmit: {\em Chaotic
  billiards generated by arithmetic groups\/}, Phys. Rev. Lett. {\bf 69} (1992)
  ~1477--1480.

\bibitem{BolSteSte92}
J.~Bolte, G.~Steil and F.~Steiner: {\em Arithmetical chaos and violation of
  universality in energy level statistics\/}, Phys. Rev. Lett. {\bf 69} (1992)
  ~2188--2191.

\bibitem{Bol93}
J.~Bolte: {\em Some studies on arithmetical chaos in classical and quantum
  mechanics\/}, Int. J. Mod. Phys B {\bf 7} (1993) ~4451--4553.

\bibitem{Sar95}
P.~Sarnak: {\em Arithmetic quantum chaos\/}, Israel Math. Conf. Proc. {\bf 8}
  (1995) ~183--236.

\bibitem{HanBer80}
J.~H. Hannay and M.~V. Berry: {\em Quantization of linear maps on a torus ---
  Fresnel diffraction by periodic grating\/}, Physica D {\bf 1} (1980)
  ~267--290.

\bibitem{Kea91c}
J.~P. Keating: {\em The cat maps: Quantum mechanics and classical motion\/},
  Nonlinearity {\bf 4} (1991) ~309--341.

\bibitem{Sar97}
P.~Sarnak: {\em Values at integers of binary quadratic forms\/}, in {\it
  Harmonic analysis and number theory} (Montreal 1996), CMS Conf. Proc. {\bf
  21}, American Mathematical Society, Providence, RI  (1997) ~181--203.

\bibitem{EskMarMoz98:p}
A.~Eskin, G.~A. Margulis and S.~Mozes: {\em Quadratic forms of signature (2,2)
  and eigenvalue spacings on rectangular 2-tori\/}, preprint  (1998).

\bibitem{PanBohGia89}
A.~Pandey, O.~Bohigas and M.-J. Giannoni: {\em Level repulsion in the spectrum
  of two--dimensional harmonic oscillators\/}, J. Phys. A {\bf 22} (1989)
  ~4083--4088.

\bibitem{Ble90}
P.~M. Bleher: {\em The energy level spacing for two harmonic oscillators with
  golden mean ratio of frequencies\/}, J. Statist. Phys. {\bf 61} (1990)
  ~869--876.

\bibitem{Ble91}
P.~M. Bleher: {\em The energy level spacing for two harmonic oscillators with
  generic ratio of frequencies\/}, J. Statist. Phys. {\bf 63} (1991) ~261--283.

\bibitem{Gre96}
C.~D. Greenman: {\em The generic spacing distribution of the two-dimensional
  harmonic oscillator\/}, J. Phys. A {\bf 29} (1996) ~4065--4081.

\bibitem{Mar98:p}
J.~Marklof: {\em The $n$-point correlations between
  values of a linear form\/}, preprint
  IHES/M/98/66, with an appendix {\it The number of solutions of simultaneous
  quadratic equations} by Z. Rudnick  (1998).

\bibitem{BerBalTabVor79}
M.~V. Berry, N.~L. Balazs, M.~Tabor and A.~Voros: {\em Quantum maps\/}, Annals
  of Physics {\bf 122} (1979) ~26--63.

\bibitem{BouBie96}
A.~Bouzouina and S.-D. Bi\`evre: {\em Equipartition of the eigenfunctions of
  quantized ergodic maps on the torus\/}, Commun. Math. Phys. {\bf 178} (1996)
  ~83--105.

\bibitem{DeBDegGia96}
S.~{De Bi\`evre}, M.~{Degli Esposti} and R.~Giachetti: {\em Quantization of a
  class of piecewise affine transformations on the torus.\/}, Commun. Math.
  Phys. {\bf 176} (1996) ~73--94.

\bibitem{MarRud99:p}
J.~Marklof and Z.~Rudnick: {\em Quantum unique ergodicity for parabolic
  maps\/}, preprint IHES/M/99/01, math-ph/9901001  (1999).

\bibitem{CorFomSin82}
I.~P. Cornfeld, S.~V. Fomin and {Ya. G. Sinai}: {\em Ergodic Theory\/}, no. 245
  in Grundlehren der Mathematischen Wissenschaften, Springer Verlag, New York,
  (1982).

\bibitem{Fur61}
H.~Furstenberg: {\em Strict ergodicity and transformation of the torus\/},
  Amer. J. Math. {\bf 83} (1961) ~573--601.

\bibitem{HarWri79}
G.~H. Hardy and E.~M. Wright: {\em An Introduction to the Theory of Numbers\/},
  Clarendon Press, Oxford, 5th edn.,  (1979).

\end{thebibliography}
\end{document}